# SEQUENTIAL MINING: PATTERNS AND ALGORITHMS ANALYSIS

*Thabet Slimani[1], and Amor Lazzez[2]*

[1] *Computer Science, Taif University & LARODEC Lab, Saudia Arabia,* [2] *Computer Science, Taif University, Saudia Arabia*

[1] *thabet.slimani@gmail.com,* [2] *a.lazzez@gmail.com*

**Abstract**

*This paper presents and analysis the common existing sequential pattern mining algorithms. It presents a classifying study of sequential pattern-mining algorithms into five extensive classes. First, on the basis of Apriori-based algorithm, second on Breadth First Search-based strategy, third on Depth First Search strategy, fourth on sequential closed-pattern algorithm and five on the basis of incremental pattern mining algorithms. At the end, a comparative analysis is done on the basis of important key features supported by various algorithms. This study gives an enhancement in the understanding of the approaches of sequential pattern mining.*

## I. INTRODUCTION

Sequential pattern is a set of itemsets structured in sequence database which occurs sequentially with a specific order. A sequence database is a set of ordered elements or events, stored with or without a concrete notion of time. Each itemset contains a set of items which include the same transaction-time value. While association rules indicate intra-transaction relationships, sequential patterns represent the correlation between transactions. Sequential pattern mining (SPM) [1] is the process that extracts certain sequential patterns whose support exceeds a predefined minimal support threshold. Additionally, sequential pattern mining helps to extract the sequences which reflect the most frequent behaviors in the sequence database, which in turn can be interpreted as domain knowledge for several purposes. To reduce the very large number of sequences into the most interesting sequential patterns and to meet the different user requirements, it is important to use a minimum support which prunes the sequential pattern with no interest. It is clear that a higher support of a sequential pattern is preferred for more interesting sequential patterns. Sequential pattern mining is used in several domains. SPM is used in business organizations to study customer behaviors. Additionally, SPM is used in computational biology to analyze the amino acid mutation patterns. SPM is also used in the area of web usage mining to mine several web logs distributed on multiple servers.

Recently, several algorithms for SPM have been proposed and most of the essential and prior algorithms are based on the property of the Apriori algorithm proposed by Agrawal and Srikant in 1994 [2]. The property states that a frequent pattern contains sub-patterns that are in turn frequent. Based on this assumption, a succession of algorithms has been proposed: in 1995, the algorithms AprioriAll, AprioriSome, DynamicSome have been proposed by Agrawal and Srikant [1]. Additionally, the Apriori-based horizontal formatting method (GSP) have been presented in 1996 by the same previous authors [3] and the Apriori-based vertical formatting method (SPADE algorithm) has been presented by Zaki en 2001 [4]. More recently, a list of algorithms based on data projection have been proposed. Among those algorithms: FreeSpan proposed by Han et al. In 2000 [4], PrefixSpan (Projection-based pattern growth method) proposed by Pei et al. In 2001[5], SPAM (Apriori-based candidate generation and pruning) proposed by Ayres et al. In 2002[6]. CloSpan proposed by Yan et al. In 2003 [7] and TSP proposed by Tzvetkov et al. In 2005 [8]. Most of these algorithms would be explained later.

## II. NOTATIONS AND CATEGORIES OF PATTERNS

As stated above, a sequence database is a set of ordered sequences. A sequence is an ordered list of events, denoted $< e_1\ e_2\ \ldots\ e_l >$. Given two sequences $\alpha = < x_1\ x_2\ \ldots\ x_n >$ and $\beta = < y_1\ y_2\ \ldots\ y_m >$, then $\alpha$ is called a subsequence of $\beta$, denoted as $\alpha \subseteq \beta$, if there exist integers $1 \leq j_1 < j_2 < \ldots < j_n \leq m$ such that $x_1 \subseteq y_{j1},\ x_2 \subseteq y_{j2},\ \ldots,$ and $x_n \subseteq y_{jn}$. If $\alpha$ and β have the following sequences α=<(xy), t> and β=< (xyz), (zt)>, then β is a super sequence of α. Given a sequence database recorded in the Table 1, it is possible to find the complete set of frequent subsequence.

Table 1: Sequence database

| SID | Sequence |
|---|---|
| 10 | <l(**lm**n)(l**n**)o(nq)> |
| 20 | <(lo)n(mn)(lp)> |
| 30 | <(pq)(**lm**)(oq)**n**m> |
| 40 | <pr(lq)nmo> |

Each row in the Table 1 is a sequence. A sequence contains a set of elements. Each element may contain an item or a set of items (marked between parenthesis). Items within an element are unordered and they listed alphabetically. As example, <l(mn)on> is a subsequence of <l(lmn)(ln)o(nq)>. Let be a support threshold minimum support equal to 2, <(lm)n> is a valid *sequential pattern*. A huge number of possible sequential patterns are hidden in databases. In sequential pattern mining problems, three main categories of patterns can be stated: Periodic or regular patterns, statistical patterns and approximate patterns [9]. We describe in the following sections all the pattern categories.

## *Periodic Patterns*

This model is used to predict the occurrences of some event (included in the data set) in the future and understand the intrinsic characteristics included in it. This model is frequently too restrictive, because if some of its occurrences are misaligned it fails to detect some interesting patterns. To enable a better flexible model, a pattern can be partially filled. The main purpose is to find the subsequences that show the periodicity in the input sequence. As an example, if we have as input sequences ({x}{y}{z}{x}{y}{z}{x}{y}{z}), the pattern {x}{y}{z} is a periodic pattern because it is repeated with a period equal to three. This previous pattern is called full periodic pattern because each position in the pattern shows the periodicity. The full periodic pattern is not available only in some applications, but many applications impose a data sequence where not every position shows the periodic pattern. As an example, if we have an input pattern like this ({x}{y}{z}{y}{x}{z}{x}{y}{x}{x}{z}{y}), then we find a novel pattern {x}*{y} where * is a wide range of items and there is no full periodic pattern with length 3. This is called a partial periodic pattern.

## *Statistically Significant Patterns*

The most used measures used to evaluate sequential patterns are the support and confidence. These measures are not meaningful for all applications of sequential pattern mining. Significant or rare patterns are the patterns missed if we use the number of occurrences (standardized support) as a measure of importance. This problem has been explored by various data mining applications. In some applications, users are interested by the k most significant patterns, and this task can be easily realized by using a threshold value and the top k patterns that have an information gain greater than the specified threshold should be returned. But, the problem of the information gain value is the difficulty to identify the location of the occurrences of the patterns. As an example: if we have tow input patterns sequences as follows S1=({x}{y}{z}{y}{x}{y}{t}{z}{x}{y}{y}{t}) and S2=({y}{z}{t}{y}{x}{y}{x}{y}{x}{y}{t}{z}), then the pattern {x}{y} has the same information gain in the two sequences, it is dispersed in S1 buts repeats consecutive in S2. Several works have been presented in [9], for more explanation.

## *Approximate Patterns*

Noisy itemsets or imperfect data are mined in a similar manner as perfect itemsets. Imperfect data occurs in some applications if the occurrence of a pattern cannot be recognized. The two previous described patterns take into account only exact match of the pattern in data. An approximate pattern is a sequence of symbols which occurs with a value greater than an approximate threshold in the data sequence. To solve the problem of approximate pattern discovery, the authors in [9] have proposed the concept of computability matrix. This matrix gives a probabilistic connection from the observed values to the real values. This computability matrix method offers the possibility to compute the true support of patterns. As an example, if we have a compatibility matrix as in Table 2:

Table 2: Compatibility matrix example

| Real values | Observed values | | | |
|---|---|---|---|---|
| | x | y | z | t |
| x | 0.9 | 0.12 | 0.0 | 0.05 |
| y | 0.11 | 0.7 | 0.1 | 0.11 |
| z | 0.0 | 0.0 | 0.88 | 0.12 |
| t | 0.1 | 0.17 | 0.0 | 0.78 |

The observed item t corresponds to a true occurrence of x, y, z and t with probability P(x,t)=0.05, P(y,t)=0.11, P(z,t)=0.12 and P(t,t)=0.78, respectively. The definition of compatibility matrix can be done by the domain expert, but there exist some methods to obtain the reasonable value of each entry in the matrix with a certain degree of error.

## II. CATEGORIES OF SEQUENTIAL PATTERN MINING ALGORITHMS

We classify the algorithms for sequential pattern mining can into the following classes: Apriori-like algorithms, BFS (Breadth First Search)-based algorithms, DFS (Depth First Search)-based algorithms, closed sequential pattern based algorithms, and incremental-based algorithms:

### *Apriori-like algorithms*

The first introduction of classical Apriori-based sequential pattern mining algorithms was in [1]. Let be a transaction database including customer sequences. This database is composed by three attributes (customer-id, transaction-time and purchased-item). The mining process was decomposed with five steps:
a) **Sort step:** which sort the transactional database according the custome-id.
b) **L-itemset step**: the objective is to obtain the large 1-itemsets from the sorted database, based on the support threshold.
c) **Transformation step**: this step replaces the sequences by those large itemsets they contain. For efficient mining, all the large itemsets are mapped into an integer series. Finally, the original database will be transformed into set of customer sequences represented by those large itemsets.
d) **Sequence step**: From the transformed sequential database, this step generates all frequent sequential patterns.
e) **Maximal step**: This step prunes the sequential patterns that are contained in other super sequential patterns, because we are only concerned with maximum sequential patterns.

Even though the Apriori algorithm is the basis of many efficient algorithms developed later, it is not efficient enough. The authors of the work in [2] have detected an interesting *downward closure* property, named Apriori, among frequent *k*itemsets: *A k-itemset is frequent only if all of its sub-itemsets are frequent.* This property means that frequent itemsets can be mined by identifying frequent 1-itemsets (first scan of the database), then the frequent 1-itemsets would be used to generate candidate frequent 2-itemsets, this process will be repeated again to obtain the frequent 2-itemsets. This process iterates until any frequent *k*-itemsets can be generated for some *k*.

There have been widespread studies on the improvements of Apriori, e.g. sampling approach [10], dynamic itemset counting [11], incremental mining [12], parallel and distributed mining [12] [13]. The work in [14], the number of candidate patterns that can be generated at the level-wise mining approach can be derived with a rigid upper bound. The obtained result reduces effectively the number of database scans.

In some cases, the size of candidate sets using the Apriori principle is significantly reduced. This situation can cause two problems:
a) A huge number of candidate sets should be generated, and b) uses of pattern matching to constantly scan the database and discovers the candidates. To encompass this problem, the work in [5] proposed an FP-growth method aiming to mine the complete set of frequent itemsets without candidate generation. FP-growth compresses the database into a frequent-pattern tree, or *FP-tree based on the* frequency-descending list. The concatenation of the suffix pattern with the frequent patterns generated from a conditional FP-tree achieves the pattern growth. Instead to find long frequent patterns, the FP-growth algorithm searches recursively for shorter suffixes and then concatenating them. This method considerably reduces search time, according performance studies. Some extensions of the FP-growth approach, including H-Mine proposed by Pei et al. In 2001 [15] which investigate a hyper-structure mining of frequent patterns; discovering the prefix-tree-structure with array-based implementation for efficient pattern growth mining by Grahne and Zhu in 2003 [16] and a pattern-growth mining with top-down and bottom-up traversal of such trees proposed in the work of Liu et al. [17][18].

### *BFS-based algorithms*

Breath-first (level-wise) search algorithms describe the Apriori-based algorithms because all *k*-sequences are constructed together in each *k*th iteration of the algorithm as they traverse the search space. Several algorithms developed using the principle of BFS algorithms. Among them we enumerate some of them in the following sections:

- ☒ **GSP algorithm**: The GSP algorithm proposed in [3], do the same work of AprioriAll algorithm, but it doesn't require finding all the frequent itemsets first. This algorithm allows a) placing bounds on the time separation between adjacent elements in a pattern, b) allowing the items included in the pattern element to span a transaction set within a time window specified by user, c) permitting the pattern discovery in different level of a taxonomy defined by user. Additionally, GSP is designed for discovering generalized sequential patterns. The GSP algorithm makes multiple passes over sequence database as follows: 1) in the first pass, it finds the frequent sequences that have the minimum support. 2) At each pass, every data sequence is examined in order to update the occurrence number of the candidates contained in this sequence. The pseudo code of GSP algorithm is as follows:

> ✓ Obtain a sequences in form of <x> as length-1 candidates
> ✓ find $F_1$ (the set of length-1 sequential patterns), after a unique scan of database
> ✓ Let k=1;
> While $F_k$ is not empty do
>     - Form $C_{k+1}$, the set of length-(k+1) candidates from $F_k$;
>     - If $C_{k+1}$ is not empty, unique database scan, find $F_{k+1}$ (the set of length-(k+1) sequential patterns)
>     Let k=k+1;
> End While

This algorithm has the following drawbacks:
- The generation if a huge set of candidate sequences, which needs a multiple scans of the database.
- The expensive number of short patterns for the mined pattern length and for that reason this algorithm is inefficient for mining long sequential patterns.

Consequently, it is important to review the sequential pattern mining problem to discover more efficient and scalable methods which may reduce the expensive candidate generation.

☒ **MFS:** Is a modified version of GSP, proposed in [19] with the aim to reduce the I/O cost needed by GSP. MFS computes as a first step the rough estimate of all the frequent sequences set as a suggested frequent sequence set and to maintain the set of maximal frequent sequences known previously it uses the candidate generation function of GSP. The results obtained in [19] show that MFS saves I/O cost significantly in comparison with GSP.

### DFS-based algorithms

The algorithms adopting this feature show only an ineffective pruning method and engender a great number of candidate sequences, which requires consuming a lot of memory in the early stages of mining. Several algorithms developed using the principle of DFS algorithms. Among them we enumerate some of them in the following sections:

☒ **SPADE:** This algorithm is proposed in [4] and it includes the features of a search space partitioning where the search space includes vertical database layout. The search space in SPADE is represented as a *lattice* structure and it use the notion of equivalence classes to partition it. It decomposes the original lattice into slighter sub-lattices, so that each sub-lattice can be entirely processed using either a breadth-first or depth-first search method (SPADE is also DFS-based method). The SPADE support counting of the candidate sequence method includes bitwise or logical operations. A conducted experimental results show that SPADE is about twice as fast as GSP. The reason behind this is that SPADE uses a more efficient support counting method based on the idlist structure. Additionally, SPADE shows a linear scalability with respect to the number of sequences. The Pseudo code of SPADE algorithm is as follows.

> Input:
> D  //ID-Lists of sequences
> S  //support
> Output:
> F  //Frequent sequences
> Begin
>     ✓ Determine frequent items, F1;
>     ✓ Determine frequent 2-sequences, F2;
>     ✓ Find equivalence classes ∈ for all 1-sequences $[S]_{\theta_1}$;
>     ✓ For each [S] ∈ ϵ do
>          Find frequent sequences F;
> End

As an example, let be the ID-List for sequences of length 1 illustrated in the Figure1. The support count for the item {A} is 3, but the support count for the element <{A,D}> is 2. The equivalent classes $\theta_1$ is presented in the Figure 2.

| A | | B | | C | | D | |
|---|---|---|---|---|---|---|---|
| Customer | Time | Customer | Time | Customer | Time | Customer | Time |
| $C_1$ | 10 | $C_1$ | 10 | $C_1$ | 20 | $C_1$ | 30 |
| $C_2$ | 15 | $C_1$ | 20 | $C_2$ | 15 | $C_2$ | 20 |
| $C_3$ | 15 | $C_2$ | 15 | $C_3$ | 15 | $C_3$ | 15 |

Figure. 1 ID-List as an example of SPADE Algorithm.

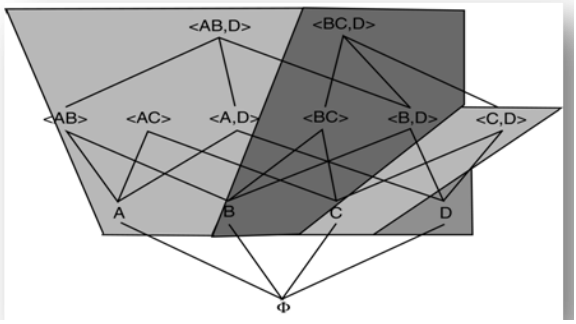

Figure. 2 The equivalent classes θ₁ generated by SPADE Algorithm

- **FreeSpan:** FreeSpan is an algorithm proposed by Pei et al. In 2001[4] with the aim to reduce the generation of candidate subsequences. It uses projected databases to generate database annotations in order to guide the mining process to rapidly find frequent patterns. The general idea of FreeSpan is to use frequent items to project sequence databases into a set of smaller projected databases recursively using the currently mined frequent sets, and subsequence fragments in each projected database are generated, respectively. Two alternatives of database projections can be used *Level-by-level* projection or *Alternative-level* projection. The method used by FreeSpan divide the data and the set of frequent patterns to be tested, and limits each test being conducted to the corresponding smaller projected database. FreeSpan scan the original database only three times, whatever the maximal length of the sequence. Experimental results show that FreeSpan is efficient and mines the complete set of patterns and it is considerably faster than the GSP algorithm. The major cost of FreeSpan is to deal with projected databases.

- **PrefixSpan:** proposed by [5], this algorithm is another form projection based algorithm. The general idea is to check only the prefix subsequences and only their corresponding postfix subsequences are projected into projected databases, rather than projecting sequence database. PrefixSpan uses a direct application of the apriori property in order to reduce candidate sequences alongside projected databases. Additionally, PrefixSpan is efficient because it mines the complete set of patterns and has a significantly faster running than both GSP algorithm and FreeSpan. The major cost of PrefixSpan, similarly to FreeSpan, is the construction of projected databases. At worst, for every sequential database, PrefixSpan needs to construct a projected database. After the database projection is done, the use of bilevel *projection* represented in FreeSpan and PrefixSpan by the *S-Matrix* [4][5] is an additional faster way to mine. The main idea of PrefixSpan algorithm (presented in the following paragraph) is to use frequent prefixes to divide the search space and to project sequence databases. Its aim is to search the relevant sequences.

**PrefixSpan($\alpha$, i, S|$\alpha$)**
**Begin**
1. Scan S|$\alpha$ once, find the set of frequent items b such that
   - b can be assembled to the last element of $\alpha$ to form a sequential pattern; or
   - <b> can be appended to $\alpha$ to form a sequential pattern.
2. For each frequent item b, appended it to $\alpha$ to form a sequential pattern $\alpha'$, and output $\alpha'$;
3. For each $\alpha'$, construct $\alpha'$-projected database S|$\alpha'$, and call PrefixSpan($\alpha'$, i+1, S|$\alpha'$).

**End**

The PrefixSpan parameters are a) $\alpha$ which represents a sequential pattern; b) $l$ is the length of $\alpha$; and c) S|$\alpha$ is the $\alpha$-projected database if $\alpha \neq <\,>$, otherwise, it is the sequence database *S*.

- **SPAM:** Proposed by [6], this algorithm uses a depth-first traversal method combined with a vertical bitmap representation to store each sequence allowing a significant bitmap compression as well as an efficient support counting. SPAM uses a vertical bitmap representation of the data which are created for each item in the dataset. Each bitmap contains a bit representing each transaction in the dataset, if item i appears in transaction j, then the bit relative to transaction j of the bitmap for item i is set to 1; otherwise it is set to 0. An efficient counting and candidate generation can be enabled if the bitmap should be partitioned aiming to make sure all transaction sequences in the database appear together in the bitmap. The bitmap representation idea of SPAM requires quite a lot of memory, so it is very efficient for those databases which have very long sequential patterns. Additionally, a significant feature of this algorithm is the outputs of new frequent itemsets in an online and incremental fashion. Experimental results show that this algorithm is more efficient compared to SPADE and PrefixSpan on large datasets, but it consumes more space compared to SPADE and PrefixSpan.

---

DFS-Pruning (node $n = (s_1, ...., s_k), S_n, I_n$)
**Begin**
(1) $S_{temp} = \varphi$.
(2) $I_{temp} = \varphi$.
(3) **For each** ($i \in S_n$)
(4)    **if** (($s_1, ....., s_k, \{i\}$) is frequent)
(5)      $S_{temp} = S_{temp} \cup \{i\}$
(6) **For each** ($i \in S_{temp}$)
(7)    DFS-Pruning(($s_1, ..............,s_k, \{i\}$),$S_{temp}$, all elements in $S_{temp}$ greater than $i$)
(8) **For each** ($i \in I_n$)
(9)    **if** (($s1, ....., sk » \{i\}$) is frequent)
(10)      $Itemp = Itemp \cup \{i\}$
(11) **For each** ($i \in I_{temp}$)
(12)    DFS-Pruning (($s1, ......, sk \cup \{i\}$), $S_{temp}$, all elements in $I_{temp}$ greater than $i$)
**End**

---

*Algorithms based on closed sequential pattern*

The algorithms of sequential pattern mining presented earlier mine the full set of frequent subsequences satisfying a minimum support threshold. Nevertheless, because a frequent long sequence contains a combined number of frequent subsequences, the mining process will generate a large number of frequent subsequences for long patterns, which is expensive in both time and space. The frequent pattern mining (itemsets and sequences) needs not mine *all* frequent patterns but the *closed* ones since it leads to a better efficiency, which can really reduce the number of frequent subsequences [7]. We present, in the following section, two recognized algorithms CloSpan and BIDE [20]:

- ☒ **CloSpan:** proposed by [7] to reduce the time and space cost when generating explosive numbers of frequent sequence patterns. CloSpan mines only frequent closed subsequences (the sequences containing no super sequence with the same support), instead of mining the complete set of frequent subsequences. The mining process used by CloSpan is divided into two stages. A candidate set is generated in the first stage which is larger than the final closed sequence set. This set is called suspicious closed sequence set (the superset of the closed sequence set). A pruning method is called in the second stage to eliminate non-closed sequences. The main difference between CloSpan and PrefixSpan is the implementation of CloSpan which are an early termination mechanism that avoids the unnecessary traversing of search space. The use of backward sub-pattern and backward super-pattern methods, some patterns will be absorbed or merged which, indeedly reduce the search space growth.

---

CloSpan (s, $D_s$, minsupp, L)
**Input**: sequence s, a projected database $D_s$, and minimum support
**Output**: the prefix search lattice L.
**Begin**
(1) Check whether a discovered sequence s' exists such that either s⊆s' or s'⊆s, and database size $L(D_s)=L(D_{s'})$
(2) **If** such super-pattern or sub-pattern exists **then**
(3) Modify the link in L**; Return**
(4) **else** insert s in L;
(5) scan $D_s$ once, find the set of frequent itemset α such that α can be appended to form a sequential pattern s◊α.
(6) If no valid α available then
**(7) Return**
**(8) For each** valid α **do**
   Call CloSpan(s◊α, $D_s$◊α, minsupp, L)
**(9) Return**

---

- ☒ **BIDE:** Proposed by Wang and Han[20] which mines closed sequential patterns without candidate maintenance by adopting a closure checking scheme, called BI-Directional Extension. BIDE avoids the problem of the *candidate maintenance-and-test* paradigm used by CloSpan. It prunes totally the search space and checks efficiently the pattern closure which consumes a much less memory in contrast to the previously developed closed pattern mining algorithms. BIDE has a linear scalability with regards to the number of sequences in the database. Nevertheless, it will lose some all-frequent-sequence mining algorithms with a high support threshold, like other closed sequence mining algorithms. Experimental results conducted in [20] show that BIDE is more efficient than CloSpan.

*Incremental-based Algorithms*

In sequential pattern mining, incremental algorithm can be used for the mining of frequent and incremental database updates (insertions and deletions). We distinguish two cases to develop an incremental algorithm: (a) The complete sequences (sequence model) are inserted into and/or removed from the original database; (b) The original database contains a sequence which is updated by appending new transactions at the end.

- **SuffixTree:** SuffixTree techniques were proposed in [21] which deal with incremental sequential pattern updating. SuffixTree has only to maintain the data reading after the update, for this reason it is a very appropriate method for incremental sequence extraction. But, this algorithm presents the complexity in space which depends on the size of the database, which presents the main limitations of this method. Additionally, the sensitivity of the position to the update operation makes *SuffixTree* very expensive for dynamic strings.

- **FASTUP:** This algorithm is proposed in [22] which presents an improvement of the candidate generation and support counting of GSP algorithm. This algorithm uses the generating-pruning method to generate and validate candidates based on the previous mining result. Performance study shows that the performance of this algorithm is better than previous algorithms for the maintenance of sequential patterns in term of speediness. Nevertheless FASTUP, includes the same limitations as GSP.

- **ISM:** This work proposed by [23] which deals with incremental sequence mining for vertical database based on the SPADE approach of sequential pattern mining. ISM assumes the availability of all the frequent sequences with their support counts and those sequences in the negative border and their support (contained in the old database) in a lattice. Additionally, ISM prunes the search space for potential new sequences based on the construction of Incremental Sequence Lattice (ISL) and the exploration of its properties. Performance study shows that ISM is an improvement in execution time by up to several orders of magnitude in practice, both for handling increments of the database, in addition to the handling interactive queries, compared with SPADE.

- **ISE**: This algorithm was proposed in [24]. ISE considers both the appending of sequences and inserting of new sequences, in contrast to ISM which only considers sequence appending. If sequence appending is considered, all the previous frequent sequences are still frequent, but if we insert new sequences, some of them may become infrequent with the same minimum support. The incremental sequential mining presented in [24] is defined as following: Let S be the original database, s is considered as incremental database where new customer sequences and transactions are inserted. $[L.sup.S]$ indicates the set of frequent sequences in S, the incremental sequential pattern mining problem is to find the frequent sequences in U=$[S.sup.s]$ with respect to the same minimum support. We expect that the length of maximal frequent sequence in the old database is l, ISE algorithm decomposes the mining into two sub problems, for those candidate sequences having a length greater than l, the GSP algorithm is used directly. An empirical evaluation conducted in [24] indicates that ISE was so efficient that it was quicker to extract an increment and to mine sequential patterns from the original database than to use the GSP algorithm.

- **GSP+ and MFS+:** GSP+ and MF+ are two algorithms proposed in [25] used to mine incremental sequential patterns based on the inserted or deleted sequences from the original database: the first are based on GSP and the second are based on MFS. Based on the set of frequent sequences obtained from mining the old database, GSP+ and MFS+ can be used to efficiently compute the updated set of frequent sequences. Performance studies show the effectiveness of GSP+ and MFS+ in term of CPU costs reduction with only a small or even negative expense on I/O cost.

- **IncSP:** Proposed in [26] as an efficient incremental updating algorithm used for sequential patterns maintenance after a nontrivial number of data sequences are added to the sequence database. IncSP uses the previously computed frequent sequences as knowledge, prunes candidates early after a process of data sequences merging, and counts supports of the sequences in to the original database and the newly appended database separately. To support again the increment database in order to accelerate the whole process, InsSP uses the candidate pruning after updating pattern.

Additionally, it uses correctly combined data sequences while preserving previous knowledge useful for incremental updating based on implicit merging. Experimental results, shows that IncSP outperforms GSP based on different ratios of the increment database to the original database excluding the situation when the increment database becomes larger than the original database.

- ☒ **IncSpan:** IncSpan is an algorithm proposed in [27] used for incremental mining over multiple database increments. Inspan algorithm development is based on two novel ideas. The first idea which presents a several good properties and lead to efficient practices is the use of a set of "*almost frequent*" sequences as the candidates in the updated database. The second idea is constituted by two optimization techniques designed to improve the performance, which are *reverse pattern matching* and *shared projection*. The first technique is used for matching a sequential pattern in a sequence. Reverse pattern matching can prune additional search space, while the appended transactions are at the end of a sequence. Shared projection is intended to reduce the number of database projections for some sequences having a common prefix. Empirical study shows that IncSpan is better than ISM and PrefixSpan on incrementally updated databases by a wide scope.

- ☒ **IncSpan+:** IncSpan+ proposed by [28] to improve IncSpan. The authors agree that the algorithm IncSpan cannot find the complete set of frequent sequential patterns in the updated database D', i.e., it violates the correctness condition. The proposed algorithm ensures the correctness of mining result in the updated database. IncSpan+ ensures two important tasks: 1) the discovery of the complete FS', which guarantees the correctness of the mining result and 2) the discovery of the complete SFS', which is helpful in incrementally maintaining the frequent patterns for further database updates.

- ☒ **MILE:** MILE proposed by [29] is an efficient algorithm to facilitate the mining process in multiple sequences. It uses recursively the knowledge of existing patterns to avoid redundant data scanning which can effectively speedup the process of new pattern's discovery. Additionally, to improve the performance of the mining process, MILE has the unique feature that can incorporate prior knowledge of the data distribution in time sequence. Empirical experiments show that MILE is significantly faster than PrefixSpan.

## III. A COMPARATIVE STUDY OF SEQUENTIAL PATTERN MINING ALGORITHMS

Depending on the management of the corresponding database, sequential pattern mining can be divided into three categories of databases, namely: a) Static database, b) Incremental database and c) Progressive database. Table 1 gives a comparison of the previously described algorithms based on the following features:

- ☒ **Database Multi-Scan:** This feature includes the original database scanning to discover whether a long list of produced candidate sequences is frequent or not.

- ☒ **Candidate Sequence Pruning**: This feature allows some algorithms (Pattern-growth algorithms, and later early-pruning algorithms) to utilize a data structure allowing them to prune candidate sequences early in the mining process.

- ☒ **Search Space Partitioning**: This feature is a characteristic feature of pattern-growth algorithms. It permits the partitioning of the generated search space of large candidate sequences for efficient memory management.

- ☒ **DFS based approach:** With the use of DFS search approach, all sub-arrangements on a path must be explored before moving to the next one.

- ☒ **BFS based approach:** This feature allows level-by-level search to be conducted to find the complete set of patterns (All the children of a node are processed before moving to the next level)

- ☒ **Regular expression constraint:** This feature has a good property called growth-based anti-monotonic. A constraint is growth-based anti-monotonic if it includes the following property: A sequence must be reachable by growing from any component which matches part of the regular expression, if it satisfies the constraint.

- ☒ **Top-down search**: This feature has the following characteristic: the mining of sequential patterns subsets can be done by the corresponding set construction of projected databases and mining each recursively from top to bottom.

- ☒ **Bottom-up search:** The Apriori-based approaches use a bottom-up search (from bottom to top), specifying every single frequent sequence.
- ☒ **Tree-projection**: This feature allows simulating split-and-project by employing conditional search on the search space represented by a tree. It is used as an alternative in-memory database because it supports counting avoidance.
- ☒ **Suffix growth vs Prefix growth**: This feature allows that the frequent subsequences exist by growing a frequent prefix/suffix; since it is usually common among a good number of these sequences. This characteristic reduces the amount of memory required to store all the different candidate sequences sharing the same prefix/suffix.
- ☒ **Database vertical projection:** Algorithms using this feature visit the sequence database only once or twice to obtain a vertical layout of the database rather than the usual horizontal form, based on the bitmap or position indication table constructed for each frequent item.

## IV. CONCLUSION

Although the concept of Sequence Data Mining is new has achieved considerable advancement in few times. Several approaches concerned by sequential pattern mining have been proposed to deal with the efficiency of the algorithms improvement either with new structures, new approaches or by the database management in the computer memory. Consequently, based on the proposed criteria's, this paper classify sequential pattern mining into five major classes (among other classes), Apriori, DFS, BFS, closed sequential pattern, and incremental pattern based algorithms. Additionally, this paper presents a comparative analysis of some mining algorithms selected form the previously described algorithms based on some features defined in the previous section.

Table 3: Comparative Study of some Sequential Pattern Mining Algorithms

|  | Apriori All | GSP | SPADE | FreeSpan | PrefixSpan | SPAM | ISM | IncSp | ISE | IncSpan | MILE |
|---|---|---|---|---|---|---|---|---|---|---|---|
| **Statical database** | √ | √ | √ | √ | √ | √ |  |  |  |  |  |
| **Incremental database** |  |  |  |  |  |  | √ | √ | √ | √ | √ |
| **DataBase MultiScan** | √ | √ |  |  |  |  |  |  | √ |  |  |
| **Candidate Sequence Pruning** |  | √ | √ |  | √ |  | √ | √ | √ | √ |  |
| **Search Space Partitioning** | √ |  |  |  |  |  |  | √ |  | √ |  |
| **DFS based approach** |  |  | √ | √ | √ | √ |  |  |  |  |  |
| **BFS based approach** |  | √ |  |  |  |  |  |  |  |  |  |
| **Regular expression constraint** |  |  |  | √ | √ |  |  |  |  |  |  |
| **Top-down search** |  |  |  | √ | √ |  |  |  |  |  |  |
| **Bottom-up search** |  | √ | √ |  |  | √ |  | √ |  |  |  |
| **Tree-projection** |  |  |  |  |  |  |  | √ |  |  | √ |
| **Suffix growth** |  |  |  |  |  |  |  |  |  |  | √ |
| **Prefix growth** |  |  |  |  | √ |  |  |  |  | √ | √ |
| **Database vertical projection** |  |  | √ |  |  | √ |  |  |  | √ |  |